\documentclass[aps, prd, showpacs, nofootinbib, superscriptaddress]{revtex4} %twocolumn

\usepackage{amsmath}
\usepackage{graphicx}
\usepackage{hhline}
\usepackage[colorlinks=true, linkcolor=blue, anchorcolor=blue, citecolor=blue, urlcolor=blue]{hyperref}
\usepackage{lipsum}
\usepackage{multirow}
\usepackage{newtxmath}
\usepackage{pdflscape}
\usepackage{verbatim}
\usepackage{xcolor}
\usepackage{scalerel} % Change the size of subscribt
\definecolor{dark green}{rgb}{0.00, 0.39, 0.00}
\usepackage{tikz} %% tikz 
\usetikzlibrary{shapes.arrows}
\usetikzlibrary{arrows.meta}
\usepackage{epsfig} %multi-row figures

\def\NT{\text{NT}}

\def\cB{\text{B}}

\def\Pl{\text{Pl}}

\def\cG{\text{G}}
\def\LPl{\Lambda_{\text{Pl}}}
\def\NAT{\text{NAT}}
\def\PAT{\text{PAT}}

\newcommand{\cw}{\columnwidth}

\def\B{\text{B}}

\def\physrep{Phys.~Rep.}	% Physics Reports
\def\plb{Phys.~Lett.~B} %Physics Letters B 
\def\grg{ Gen.~Rel.~Grav.} 
\def\ijmpd{Int.\ J.\ Mod.\ Phys.\ D}

\makeatletter
\def\url@rmstyle{%
  \@ifundefined{selectfont}{\def\UrlFont{\sf}}{\def\UrlFont{\footnotesize\rmfamily}}}
\makeatother
\def\jcap{JCAP} 		%
\begin{document}

\title{A solution to the initial condition problems of inflation : NATON}
%%% Authors
\author{Seokcheon Lee}\thanks{E-mail: skylee2@gmail.com}
\affiliation{Department of Physics, BK21 Physics Research Division, Institute of Basic Science \\ Sungkyunkwan University, Suwon 16419, South Korea}

\date{\today}

\begin{abstract}
The recent astonishing realization of the negative absolute temperature (NAT) for motional degrees of freedom \cite{Braun_ea13} inspires its possible application to the early universe. The existence of the upper bound on the energy of the system is the sufficient requirement for a NAT and this might correspond to the Planck scale at the Big Bang model. It has been argued that standard inflation requires an initial patch that is smooth over distance scales a bit larger than the causal horizon at the onset of inflation. We show that this initial condition problem can be solved if the NAT fermion, dubbed  ``NATON'', occupies the Universe before the standard inflation is ignited by providing mini inflation prior to the standard one. As long as NATON dominated era lasts until at least ten times older than the Planck time, it makes the Universe homogeneous enough to derive the successful standard inflation.  %It has been known that the natural initial condition is required in order to obtain a period of cosmological inflation. 

\end{abstract}

\maketitle

\section{Introduction}
\label{sec:introduction}

Current observations on the thermal history of the Universe and the growth of cosmic structures are in good agreement with the predictions of the Big Bang model despite its drawback of fine-tuning problems \cite{Tanabashi_ea2018}. Inflation models based on the scalar field(s) not only solve these initial value problems but also generate the seed for the large scale structure and anisotropies of the cosmic microwave background \cite{Tanabashi_ea2018}. 

Even though the inflationary paradigm makes a great achievement in the prediction of the current Universe, inflation should arise without requiring special initial conditions in order to be a successful model. It requires an initial patch should be smooth over distance scales a bit larger than the causal horizon at the onset of inflation. At present, the naturalness of initial conditions (ICs) for inflation is still being debated \cite{Mazenko_ea85,Linde_85, Goldwirth_92, Knox_ea93, Iguchi_ea97, Bolejko_ea10, Alho_ea10, Ijjas_ea13, Ijjas_ea14, Guth_ea14, Linde_14, Easther_ea14, Dalianis_ea15, Carrasco_ea15, Kaya_16, Clough_ea16, Brandenberger_16, Bastero-Gil_ea16, Bagchi_ea18, Linde_18, Mishra_ea18}. One of the main reasons for these IC problems stems from the assumption on the fact that the early universe after the Big Bang is dominated by radiations (or matters) before the inflation starts.     

Since Onsager adopts the concept of the negative absolute temperature (NAT) to characterize the equilibria of the turbulent flow with long-range interacting point vortices in 1949 \cite{Onsager_49}, both theoretical and experimental pioneer works using the nuclear spin systems related to NAT are followed at the 1950s \cite{Purcell_ea51, Ramsay_56}. It takes more than half a century that the first experiment of NAT in a system with motional degrees of freedom (an ultra-cold bosonic gas) is realized \cite{Braun_ea13}.

The NAT is produced when the system has more population in the high energy states compared to the low energy ones \cite{Puglisi_ea17}. In equilibrium statistical mechanics, an inverse temperature parameter (also known as coldness) $\beta = (k_{\B} T)^{-1}$ can range in the full infinite real line ($-\infty, + \infty$) without any inconsistency. The essential requirement for a NAT is the existence of an upper bound to the energy of the system in order to describe the normalizable matter distribution. Either an absolute upper bound or an energy gap allowing a metastable population inversion can act as the upper bound in the condensed matter system. At the early universe, the Planck energy scale is the maximum energy scale where the current theory of space and time properly work. 

Thus, one can naturally expect that there can be a NAT dominated epoch right after the Big Bang due to this existence of an upper bound (Planck) scale. In equilibrium, the requirement of the maximum entropy principle leads to the same sign of both the absolute pressure and temperature ($\partial S/ \partial V |_{\text{eq}} \equiv P/T > 0$). Thus, a NAT implies a negative pressure. If the Universe is dominated by a negative pressure component, then the expansion of the Universe is accelerated. Since a NAT is hotter than all positive absolute temperatures (PAT), if it thermally contacts with the PAT components, then it will lose the heat and becomes a PAT. The duration of this process is very short in the laboratory (typically microseconds) and that is why it is not easy to find the NAT system in a daily life experience. However, the time period between the Big Bang and the inflation is $10^{-43} \sim 10^{-34}$ seconds. Thus, the typical duration time of the laboratory NAT experiments is extremely long enough compared to the epochs of the early universe.

The above description explains all the necessary ingredients to solve the IC problems of the standard inflation models. In this manuscript, we propose that the Universe is dominated by NAT fermion, dubbed ``NATON'', right after the Big Bang and it causes the mini inflationary epoch prior to the stand inflationary one. As NATON is thermally contacted with the radiation components having a positive temperature, it becomes a usual matter component ($P = 0$ as shown in the context). Thus, the Universe becomes the radiation (or matter) dominated epoch right before the beginning of the standard inflation. Compared to the traditional inflation scenario, the existence of the mini inflation due to the NATON between the Big Bang and the standard inflation makes the Universe naturally homogeneous and it can solve the IC problems of the standard inflation models.   

This manuscript is organized as follows. In the next section, we investigate the properties of NATON and explicitly show the analytic forms of its number density, energy density, and pressure. A brief review on the IC problems of inflation models is given in section \ref{sec:ICpro}. In section \ref{sec:ICsol}, the mechanism of how the existence of a NATON dominated epoch solves the IC problems is explained. We also show how long it should dominate the Universe to solve the problems. We summarize and conclude in section \ref{sec:DisCon}.      

\section{Negative Absolute Temperature}
\label{sec:NAT}
In this section, we investigate the role and the properties of a NATON in the early Universe. The existence of the finite upper limit of the system's energy might permit a system of NAT. The Big Bang model naturally provides this upper limit on the energy scale and NATON might exist between the Big Bang and the standard inflation. We assume that the NATON is a fermion and this argument is similar to that in \cite{Vieira_ea16}. However, we explicitly show the analytic forms of energy density and the pressure of a NATON. Also, the role of a NATON is producing the epoch of mini inflation prior to the standard inflation. This model has not been proposed yet.

\subsection{Maximum number density and maximum energy density of NATON}
\label{ssec:maxnrhoNATON}

\begin{figure}[h]
\centering
\vspace{1cm}
\begin{tabular}{cc}
\epsfig{file=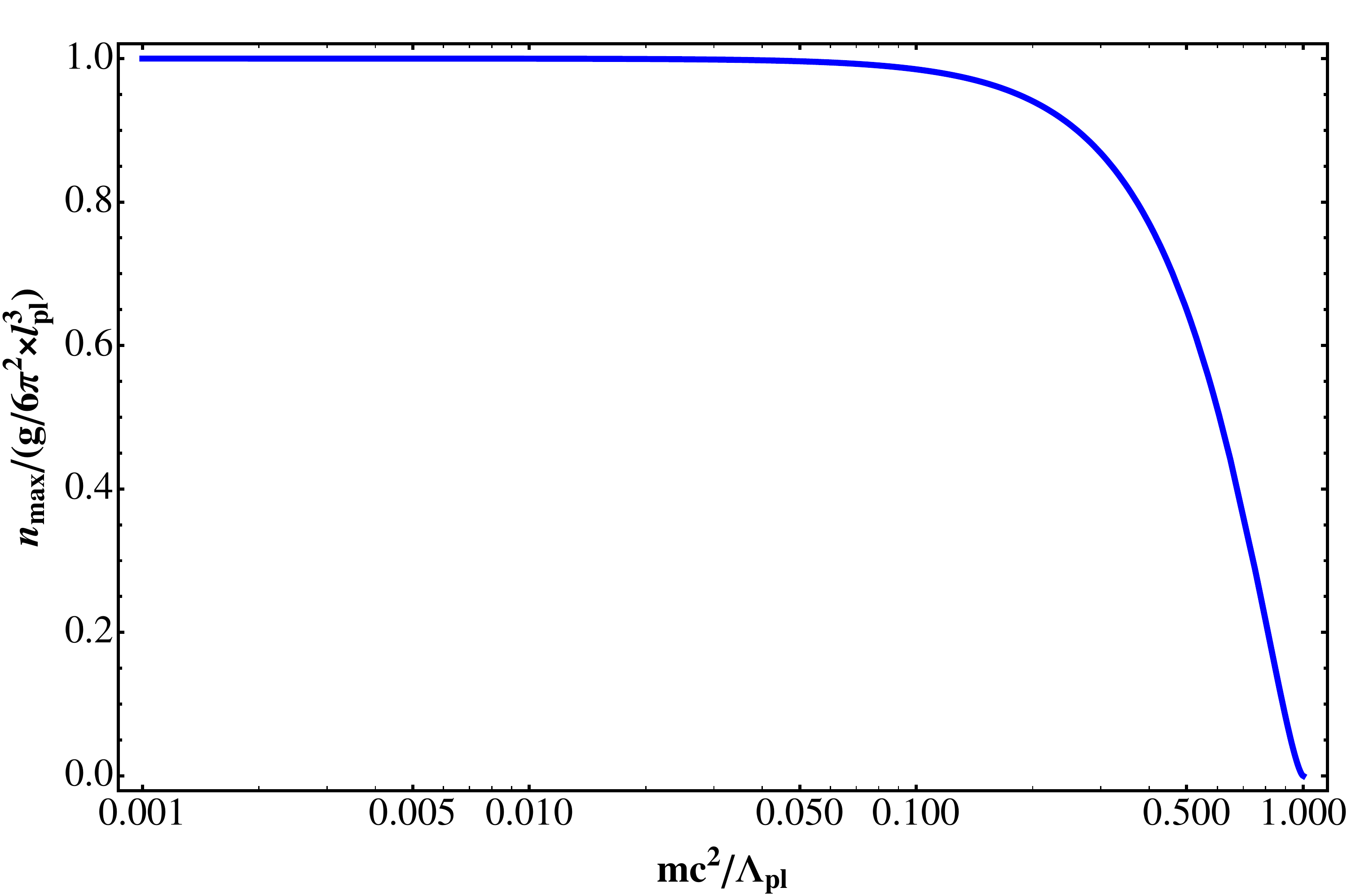,width=0.5\linewidth,clip=} &
\epsfig{file=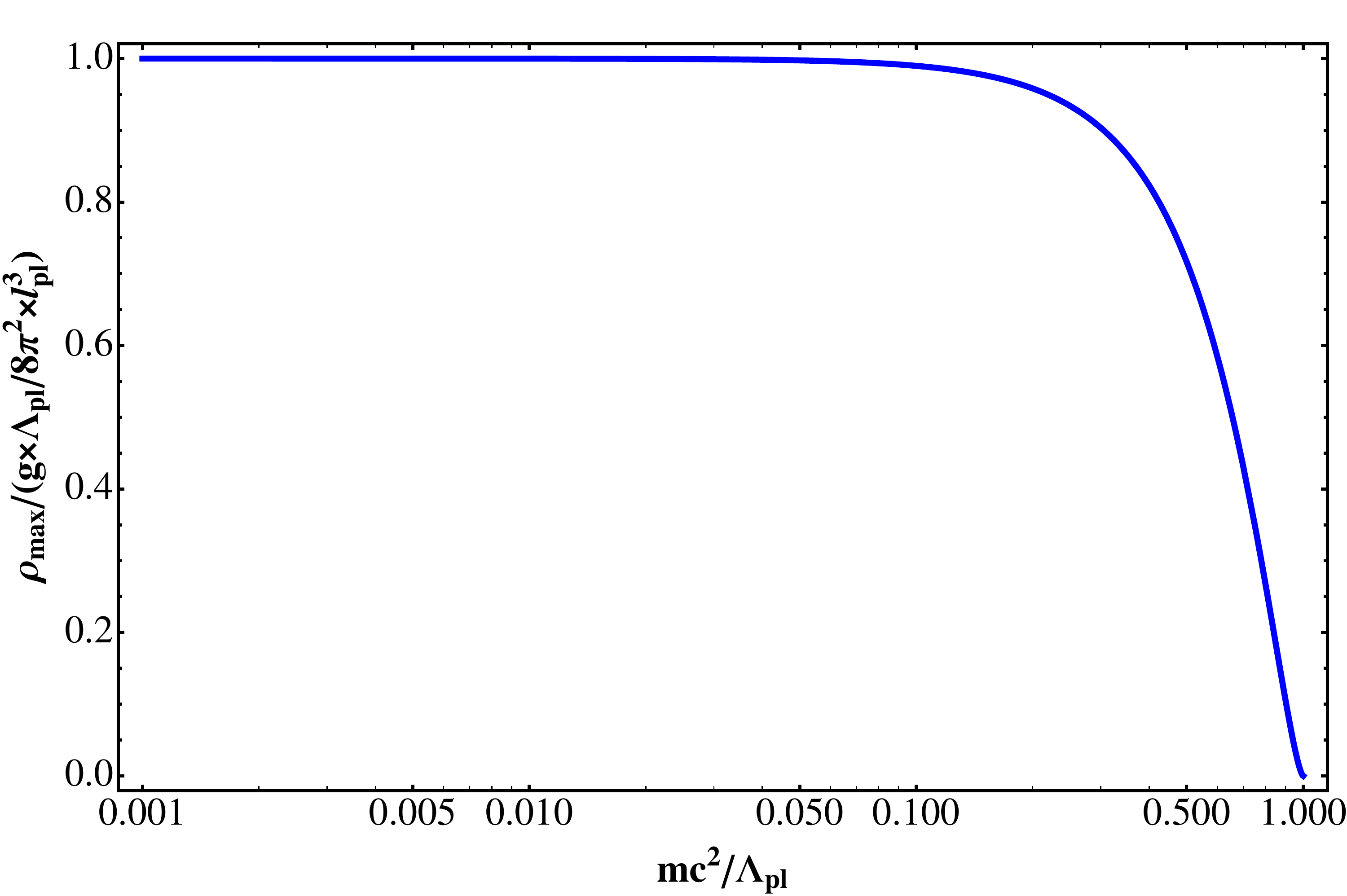,width=0.5\linewidth,clip=}
\end{tabular}
\vspace{-0.5cm}
\caption{The maximum number density $n_{\text{max}}$ and the maximum energy density $\rho_{\text{max}}$ of NATON as a function of its mass $m$. a) When the NATON mass approaches to the Planck mass, the maximum number density goes to zero. As the mass is less than 10 \% of the Planck, then $n_{\text{max}}$ saturates to its maximum value $g/(6 \pi^2 l_{\Pl}^3)$. b) The mass dependence of the maximum energy density. As $m$ increases, $\rho_{\text{max}}$ decreases. It reaches to the saturation value $g \LPl/ (8 \pi^2 l_{\Pl}^3)$ as $m$ decreases.} \label{fig-1}
\vspace{1cm}
\end{figure}

The density of states of a particle with the internal degrees of freedom $g$ is given by $g/(2\pi \hbar)^{3}$.  If one considers fermions, then one uses the phase space distribution function with Fermi-Dirac (FD) one
\begin{align} 
f(\vec{x}, \vec{p}, t ) &= f(p) = \frac{1}{e^{\beta \left( E - \mu N \right)} + 1} \label{fepsilon} \, , 
\end{align}
where we use the fact that distribution is homogeneous and isotropy to use $f(\vec{x}, \vec{p}, t ) = f(p)$. We also use the definition of the reciprocal of the thermodynamic temperature $\beta = \left( k_{\B} T \right)^{-1}$, the magnitude of the energy $E = \sqrt{|\vec{p}|^2 c^2 + m^2 c^4}$, the chemical potential $\mu$, and the total number of particle $N$. Then, the number density $n$ and the energy density $\rho$ are given by 
\begin{align}
n(\beta,\mu, m) &\equiv \frac{g}{(2\pi \hbar)^3} \int_{0}^{p_{\scaleto{\Pl\mathstrut}{5pt}}}  d^3 p f(p) =  \frac{g}{2 \pi^2 \hbar^3} \int_{0}^{p_{\scaleto{\Pl\mathstrut}{5pt}}} \frac{p^2}{\exp \left[ \beta \left( \sqrt{p^2 c^2 + m^2 c^4} - \mu N \right) \right] + 1} dp \nonumber \\ 
		  &= \frac{g}{2 \pi^2 \left( \hbar c \right)^3}\int_{mc^2}^{\LPl} \frac{E \sqrt{E^2 - m^2 c^4}}{\exp \left[ \beta \left( E - \mu N \right) \right] + 1} dE \,\, , \label{nTmu} \\
\rho(\beta,\mu, m) &\equiv \frac{g}{(2\pi \hbar)^3} \int_{0}^{p_{\scaleto{\Pl\mathstrut}{5pt}}}  d^3 p E(p) f(p)  = \frac{g}{2 \pi^2 \hbar^3} \int_{0}^{p_{\scaleto{\Pl\mathstrut}{5pt}}} \frac{p^2 \sqrt{p^2 c^2 + m^2 c^4} }{ \exp \left[ \beta \left( \sqrt{p^2 c^2 + m^2 c^4} - \mu N \right) \right] + 1} dp \nonumber \\ 
		  &= \frac{g}{2 \pi^2 \left( \hbar c \right)^3} \int_{mc^2}^{\LPl} \frac{E^2 \sqrt{E^2 - m^2 c^4}}{ \exp \left[ \beta \left( E - \mu N \right) \right] + 1} dE \, , \label{rhoTmu}
\end{align}
where $p_{\scaleto{\Pl\mathstrut}{5pt}}^{2} c^2  = \LPl^2-m^2 c^4$, $\LPl = m_{\Pl} c^2$ is the Planck energy,  and $m$ is the mass of a NATON. One  can introduce the dimensionless parameters $\xi \equiv \beta pc$, $x \equiv \beta mc^2$, and $y \equiv \beta \mu N$ to rewrite the above equations
\begin{align}
n(\beta, y, x) &\equiv \frac{g}{2\pi^2} \frac{1}{ \left(\beta \hbar c \right)^{3}} \int_{0}^{\xi_{\scaleto{\Pl\mathstrut}{5pt}}}  \frac{\xi^2}{\exp \left[ \sqrt{\xi^2 + x^2} - y\right] + 1} d\xi \, , \label{nTmu2} \\
\rho(\beta,y, x) &\equiv \frac{g}{2\pi^2} \frac{\beta^{-1}}{ \left(\beta \hbar c \right)^{3}} \int_{0}^{\xi_{\scaleto{\Pl\mathstrut}{5pt}}}  \frac{\xi^2 \sqrt{\xi^2 + x^2} }{\exp \left[ \sqrt{\xi^2 + x^2} - y\right] + 1} d\xi  \, . \label{rhoTmu2}
\end{align}
The integrals in the above equations \eqref{nTmu2} and \eqref{rhoTmu2} have to be evaluated numerically. However, as $\beta \rightarrow -\infty$ ({\it i.e.} $T \rightarrow - 0$), the FD distribution becomes $1$ and one can obtain the maximum possible values of $n$ and $\rho$ without considering the number change ({\it i.e.} $y = 0$) 
\begin{align}
n_{\text{max}} (-0, 0, m) &\equiv \frac{g}{2\pi^2} \frac{1}{ \left(\beta \hbar c \right)^{3}} \int_{0}^{\xi_{\scaleto{\Pl\mathstrut}{5pt}}} \xi^2 d\xi = \frac{g}{6 \pi^2} \left( \frac{\LPl}{\hbar c} \right)^{3} \left(1 - \frac{m^2 c^4}{\LPl^2} \right)^{3/2} \nonumber \\ 
					    &\equiv \frac{g}{6 \pi^2} \frac{1}{l_{\Pl}^3} \left(1 - \frac{m^2 c^4}{\LPl^2} \right)^{3/2}\, , \label{nTmumax} \\ 
%&= \frac{g}{6 \pi^2} \left( \frac{mc}{\hbar} \right)^{3} \left(\frac{\LPl^2}{m^2 c^4} - 1 \right)^{3/2} 
%\equiv \frac{g}{6 \pi^2} \frac{1}{l_{\Pl}^3} \left(1 - \frac{m^2 c^4}{\LPl^2} \right)^{3/2} \nonumber \\
\rho_{\text{max}} (-0, 0, m) &\equiv \frac{g}{2\pi^2} \frac{\beta^{-1}}{ \left(\beta \hbar c \right)^{3}} \int_{0}^{\xi_{\scaleto{\Pl\mathstrut}{5pt}}}  \xi^{3} d\xi = \frac{g}{8\pi^2} \left( \frac{\LPl}{\hbar c}\right)^{3} \LPl \left(  \left(1 - \frac{m^2c^4}{2\LPl^2} \right) \sqrt{1 - \frac{m^2c^4}{\LPl^2} } - \frac{m^4c^8}{2\LPl^4} \ln \left[\frac{1 + \sqrt{1 - m^2 c^4 / \LPl^2}}{mc^2/\LPl}  \right]\right) \nonumber \\
						   &\equiv \frac{g}{8\pi^2} \frac{\LPl }{l_{\Pl}^3} \left( \left(1 - \frac{m^2c^4}{2\LPl^2} \right) \sqrt{1 - \frac{m^2c^4}{\LPl^2} } - \frac{m^4c^8}{2\LPl^4} \ln \left[\frac{1 + \sqrt{1 - m^2 c^4 / \LPl^2}}{mc^2/\LPl}  \right] \right) \, , \label{rhoTmumax} 
%\int_{mc^2}^{\LPl} \frac{g}{(2\pi \hbar)^3} E d^3 p = \frac{g \left(\hbar c \right)^{-3}}{2\pi^2} \int_{mc^2}^{\LPl} \sqrt{E^2 - m^2 c^4} E^2 dE \, , \label{rhoTmumax} \\
%&= \frac{g}{8\pi^2} \left( \frac{\LPl}{\hbar c}\right)^{3} \LPl \left(  \left(1 - \frac{m^2c^4}{2\LPl^2} \right) \sqrt{1 - \frac{m^2c^4}{\LPl^2} } - \frac{m^4c^8}{2\LPl^4} \ln \left[\frac{1 + \sqrt{1 - m^2 c^4 / \LPl^2}}{mc^2/\LPl}  \right]\right)  \nonumber \\
%&= \frac{g}{8\pi^2} \frac{\LPl }{l_{\Pl}^3} \left( \left(1 - \frac{m^2c^4}{2\LPl^2} \right) \sqrt{1 - \frac{m^2c^4}{\LPl^2} } - \frac{m^4c^8}{2\LPl^4} \ln \left[\frac{1 + \sqrt{1 - m^2 c^4 / \LPl^2}}{mc^2/\LPl}  \right] \right) \nonumber 
\end{align} 
where we use the definition of the Planck length $\l_{\Pl} = \hbar c / \LPl$. As shown in the Fig.\ref{fig-1}, both its maximum number density $n_{\text{max}}$ and the maximum energy density $\rho_{\text{max}}$ depend on the mass of the NATON. The values on vertical axis are normalized by their maximum values. If its mass is about the same magnitude as Planck mass, then both $n_{\text{max}}$ and $\rho_{\text{max}}$ become zero. Thus, if the mass of the NATON is close to Planck mass, then it will not contribute to the dynamics of the Universe. If the mass is one order of magnitude smaller than the Planck mass $10^{-1} m_{\Pl}$, then both values reach to their saturated maximum values, $g/(6 \pi^2 l_{\Pl}^3)$ and  $g \LPl/ (8 \pi^2 l_{\Pl}^3)$, respectively. In the left panel of this figure, the values of $n_{\max}$ for the different NATON mass are depicted. Its maximum value saturates to $n_{\text{max}}$ as $\beta$ approaches to the negative infinity. And $n_{\text{max}}$ becomes zero as the mass approaches to Planck mass. The mass dependence of $rho_{\max}$ on NATON mass $m$ are shown in the right panel of Fig.\ref{fig-1}. Its behavior is quite similar to that of $n_{\text{max}}$.  
%%%%%%%%%%%%%%%%%%%%%%%%%%%%%%%%%%%%%%%
%%%%%%%%%%%%%%%%%%%%%%%%%%%%%%%%%%%%%%%

\subsection{Energy density and pressure of NATON}
\label{ssec:rhoPNATON}

Because we focus on the early epoch of the Universe and also from the fact that the contribution of the NATON is only important when its mass is much smaller than the Planck mass, it is suitable to consider only the relativistic limits ($\beta mc^2 \rightarrow 0$) case. For the cosmological consideration, one needs to investigate the pressure of the NATON. This can be obtained from the grand potential $\Phi_{\cG}$ which is defined as \footnote{If a system contains several different types of particles then one needs to replace the last term in the first equality with $\sum_{i} \mu_{i} N_{i}$. } 
\begin{align}
\Phi_{\cG} &\equiv U - TS -\mu N \equiv - k_{\cB} T \ln {\cal Z} \equiv -\frac{1}{\beta} \ln {\cal Z} \,\, , \label{PhiG} 
\end{align}
where $U = \sum_{i} E_{i}$ is the total internal energy, $S$ is the entropy, $N = \sum_{i} N_{i}$ is the total number of particles, and ${\cal Z}$ is the partition function of the grand canonical ensemble (GCE) ${\cal Z} = \sum_{i} e^{-\beta \left(E_{i} - \mu_{i} N_{i} \right)}$. The total derivative of $\Phi_{\cG}$ is given by
\begin{align}
d \Phi_{\cG} &= dU - T dS - S dT - \mu dN - N d\mu \nonumber \\
		    &= - S dT - P dV - N d \mu \, , \label{dPhiG} \\
		    &\equiv \left( \frac{\partial \Phi_{\cG}}{\partial T} \right)_{V,\mu} dT + \left( \frac{\partial \Phi_{\cG}}{\partial V} \right)_{T,\mu} dV   + \left( \frac{\partial \Phi_{\cG}}{\partial \mu} \right)_{V, T} d\mu \nonumber 	
\end{align}
where $P$ and $V$ indicate the pressure and the volume, respectively. We use the first thermodynamics law ($dU = T dS - P dV + \mu dN$) in the second equality. Subscripts in each term denote their constant values of corresponding quantities. From the equation \eqref{dPhiG}, one can obtain the definition of $S, P$, and $N$ as
\begin{align}
S &\equiv -\left( \frac{\partial \Phi_{\cG}}{\partial T} \right)_{V,\mu} \,\, , \,\, P \equiv -\left( \frac{\partial \Phi_{\cG}}{\partial V} \right)_{T,\mu} \,\, , \,\, N \equiv - \left( \frac{\partial \Phi_{\cG}}{\partial \mu} \right)_{V, T} \, . \label{SPN} 
\end{align} 
The (Gibbs) entropy of the GCE is given by
\begin{align}
S_{\cG} &= k_{\cB} \frac{\partial}{\partial T} \left( T \ln {\cal Z} \right) \,\, . \label{SG} 
\end{align}
In equilibrium, one can replace the grand potential with the negative work ($\Phi_{\cG} = -W = -P V$) to obtain the relation between the pressure, the energy density, and the entropy density as
\begin{align}
T S &= U + P V - \mu N \,\, \rightarrow \,\, T s V = \left( \rho + P  - \mu n \right) V \nonumber \\
P &= - \rho + s T + \mu n \,\, , \label{Pther}
\end{align}
where $s$ is the entropy density. The pressure is obtained from the grand potential as defined in the equation \eqref{SPN}. For fermions, the partition function is given by
\begin{align}
{\cal Z} &= \sum_{s} e^{-\beta \left(E_{s} - \mu N_{s} \right)} = \sum_{N_i} \prod_{i} e^{-\beta \left( \epsilon_{i} - \mu \right) N_{i}} = \prod_{i} \left( 1 + e^{-\beta \left( \epsilon_{i} - \mu \right) } \right) \, , \label{calZ} 
\end{align}
where $s$ are the states of the whole system, $N_s$ means a sum over all possible combinations of $N_i$, and $i$ denotes the different one-particle states, $\epsilon_{i}$ and $N_{i}$ represents their energy and occupation number, respectively. By using the Eqs. \eqref{PhiG}, \eqref{SPN}, and \eqref{calZ} and taking the continuous limit, one obtains
\begin{align}
P(\beta,\mu, m) &= \frac{g}{2 \pi^2 } \frac{1}{(\hbar c)^{3}} \frac{1}{\beta} \int_{mc^2}^{\LPl} E^2 \ln \left[ 1 + e^{-\beta \left( E - \mu N \right)} \right] d E \label{PTmu} \, .   
\end{align}
In the relativistic limits, one can rewrite equations \eqref{nTmu2}, \eqref{rhoTmu2}, and \eqref{PTmu} as
\begin{align}
n(\beta, 0, 0) &\equiv \frac{g}{2\pi^2} \frac{1}{ \left(\beta \hbar c \right)^{3}} \int_{0}^{\xi_{\scaleto{\Pl\mathstrut}{5pt}}}  \frac{\xi^2}{e^{\xi} + 1} d\xi \, , \label{nTrel} \\
\rho(\beta, 0, 0) &\equiv \frac{g}{2\pi^2} \frac{\beta^{-1}}{ \left(\beta \hbar c \right)^{3}} \int_{0}^{\xi_{\scaleto{\Pl\mathstrut}{5pt}}}  \frac{\xi^3}{e^{\xi} + 1} d\xi  \, . \label{rhoTrel} \\
P(\beta, 0, 0) &\equiv \frac{g}{2\pi^2} \frac{\beta^{-1}}{ \left(\beta \hbar c \right)^{3}} \int_{0}^{\xi_{\scaleto{\Pl\mathstrut}{5pt}}} \xi^2 \ln \left[ 1 + e^{-\xi} \right] d\xi  \, . \label{PTrel}
\end{align}
All of the above integral are related to the (in)complete FD integral \footnote{The FD integral of order n is given by $\int_{0}^{\infty} \frac{t^{n-1}}{e^{t-x} +1} dt = - \Gamma[n] \text{Li}_{n}(-e^{x})$ where $\Gamma$ is the Euler beta function and $\text{Li}_{n}[z]$ is the polylogarithm function.} which has the solution as the polylogarithm function of a order n, $\text{Li}_{n}[z]$ \cite{Poly}. In the relativistic limits with $\mu =0$, then $n$, $\rho$, and $P$ are given by 
\begin{align}
n(\beta) &= \frac{g}{6 \pi ^2l_{\text{Pl}}^3} \frac{1}{ \beta ^3 \LPl ^3 }  \bigg(\beta ^2 \LPl ^2 \left(\beta  \LPl -3 \ln \left[1+e^{\beta  \LPl }\right]\right) -6 \beta  \LPl  \text{Li}_{2}\left[-e^{\beta  \LPl }\right] +6 \text{Li}_{3}\left[-e^{\beta  \LPl }\right]+\frac{9 }{2} \zeta[3] \bigg) \,  \nonumber \\
&\equiv n_{\text{max}}^{(0)} \bigg(1 - \frac{3\ln \left[1+e^{\beta  \LPl }\right]}{\beta \LPl} - \frac{6\text{Li}_{2}\left[-e^{\beta  \LPl }\right]}{\beta^2 \LPl^2} + \frac{6}{\beta^3 \LPl^3} \big( \text{Li}_{3}\left[-e^{\beta  \LPl }\right]  + \frac{3}{4} \zeta[3] \big) \bigg) \label{nbeta} \, , \\
\rho(\beta) &= \frac{\text{g$\LPl $} }{8 \pi ^2l_{\text{Pl}}^3} \frac{1}{ \beta ^4 \LPl ^4 } \bigg(\beta ^3 \LPl ^3 \left( \beta \LPl -4  \ln\left[1+e^{\beta  \LPl }\right] \right) -12 \beta \LPl \left( \beta \LPl  \text{Li}_{2}\left[-e^{\beta  \LPl }\right] -2 \text{Li}_{3} \left[-e^{\beta  \LPl }\right] \right) \nonumber \\ 
& \,\,\,\, -24 \text{Li}_{4} \left[-e^{\beta  \LPl }\right]-\frac{7 \pi ^4}{30} \bigg) \nonumber \\
&\equiv \rho_{\text{max}}^{(0)} \bigg( 1 - \frac{4\ln\left[1+e^{\beta  \LPl }\right]}{\beta \LPl} - \left( \text{Li}_{\text{sum}}(\beta) + \frac{7 \pi^4}{30} \right) \bigg) \,\, , \label{rhobeta} \\
P(\beta) &= \frac{g \LPl }{8 \pi ^2l_{\text{Pl}}^3} \frac{1}{\beta ^4\LPl^4} \frac{1}{3} \bigg(- 3 \beta^{4} \LPl^{4} -12 \beta \LPl \big( \beta \LPl  \text{Li}_{2}\left[-e^{\beta  \LPl }\right]  -2 \text{Li}_{3} \left[-e^{\beta  \LPl }\right] \big) -24 \text{Li}_{4} \left[-e^{\beta  \LPl }\right]-\frac{7 \pi ^4}{30} \bigg) \nonumber \\
&= - \rho_{\text{max}}^{(0)} \bigg( 1 + \frac{1}{3} \left( \text{Li}_{\text{sum}}(\beta)  + \frac{7\pi^4}{30} \right) \bigg)  \, , \,\, \text{where} \label{Pbeta} \\
\text{Li}_{\text{sum}}(\beta) &\equiv \frac{12 \text{Li}_{2} \left[ -e^{\beta  \LPl } \right] }{\beta^2 \LPl^2}  - \frac{24 \text{Li}_{3} \left[ -e^{\beta  \LPl } \right] }{\beta^3 \LPl^3}  + \frac{24 \text{Li}_{4} \left[-e^{\beta  \LPl }\right] }{\beta^4 \LPl^4} \label{Linbeta} \, , 
\end{align}
where $\zeta$ is the Riemann zeta function, $\text{Li}_{n}[z]$ is the Polylogarithm function of order n \cite{Poly}, and $n_{\text{max}}^{(0)} \left(\rho_{\text{max}}^{(0)} , P_{\text{max}}^{(0)} \right)$ denotes $n_{\text{max}}(m=0) \Big( \rho_{\text{max}}(m=0), P_{\text{max}}(m=0) \Big)$. 

%%%%%%%%%%%%%%%%%%%%%%%%%%%%%%%%%%%%%%%
\begin{figure}[h]
\centering
\vspace{1cm}
\begin{tabular}{ccc}
\epsfig{file=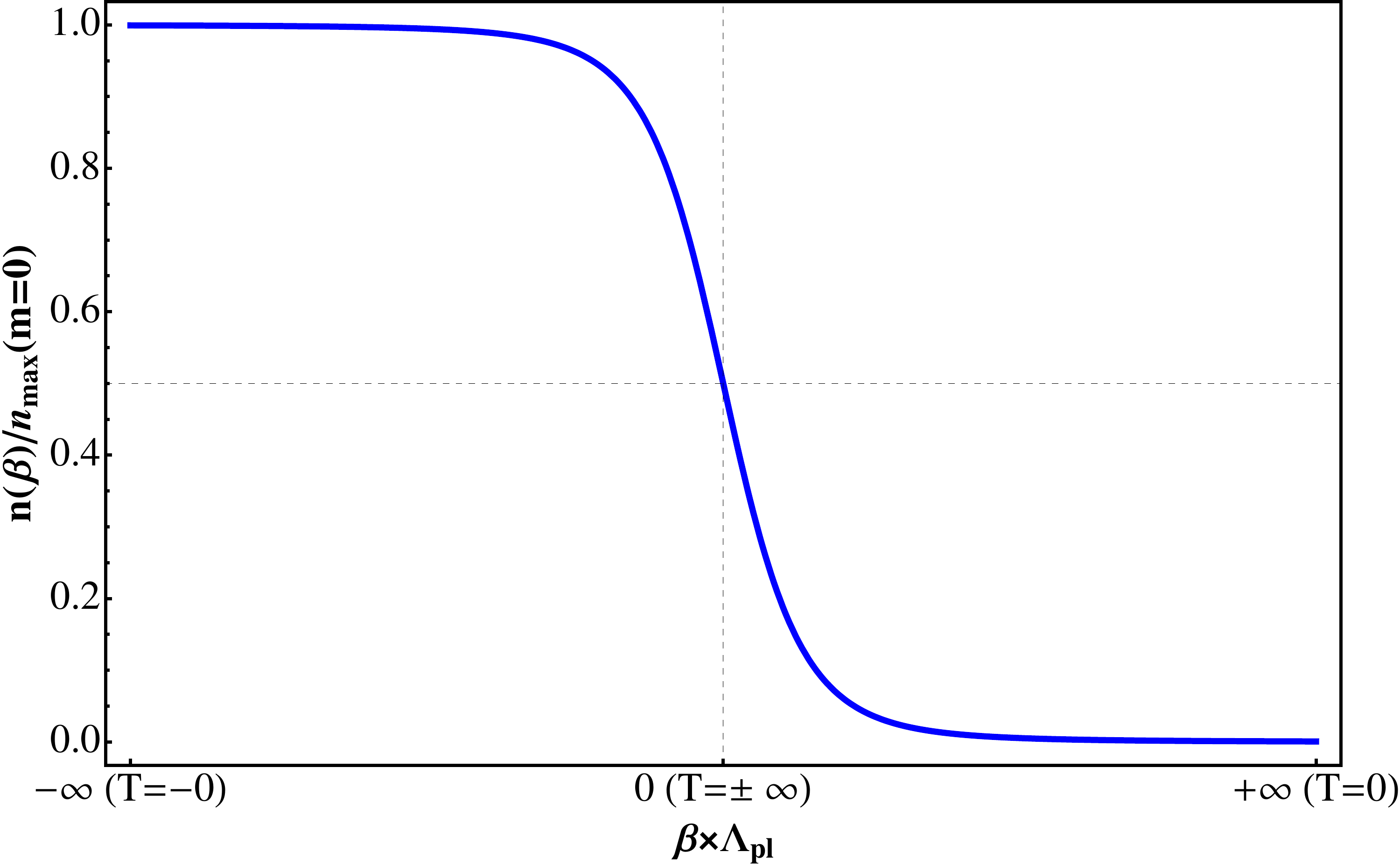,width=0.33\linewidth,clip=} &
\epsfig{file=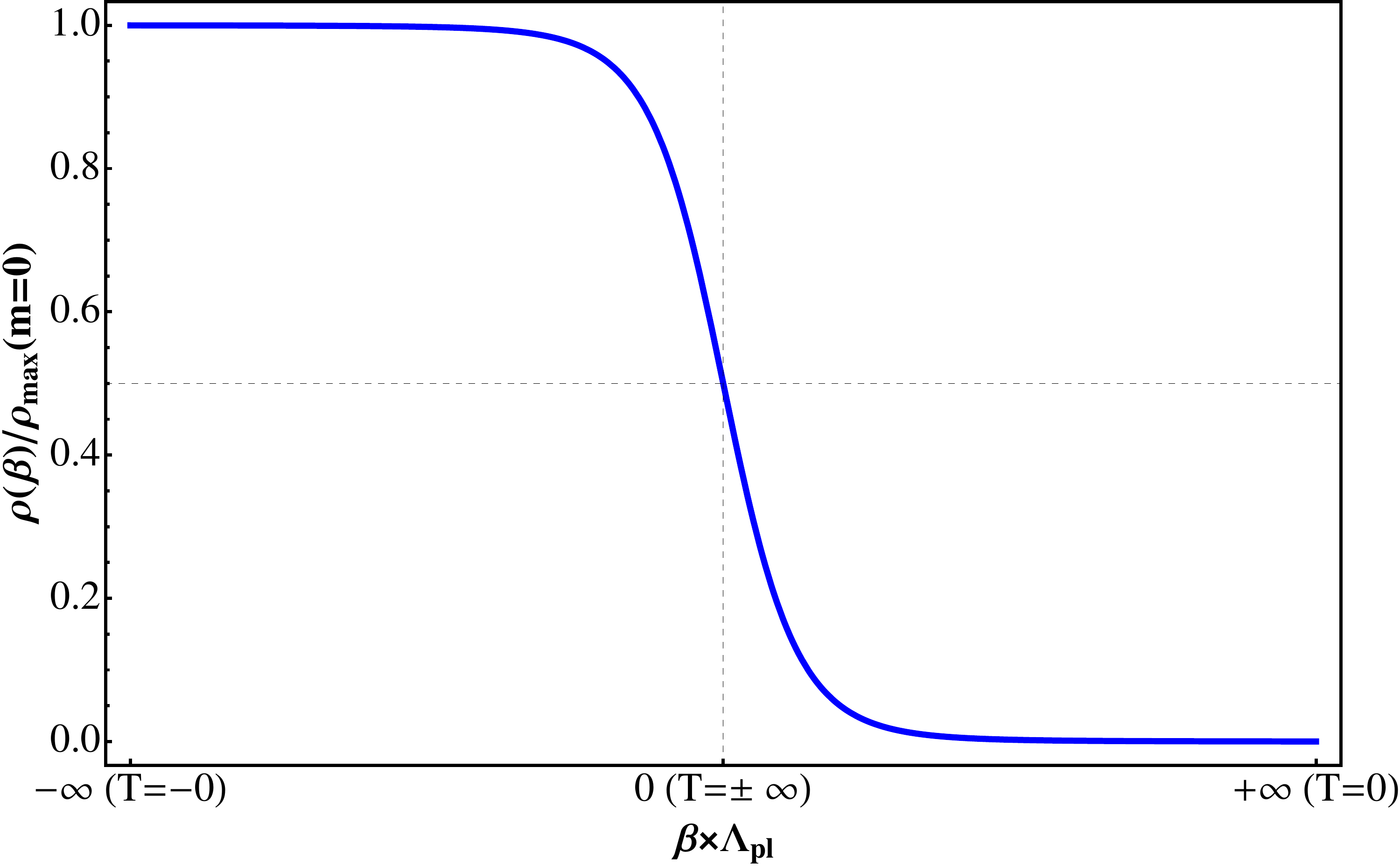,width=0.33\linewidth,clip=} &
\epsfig{file=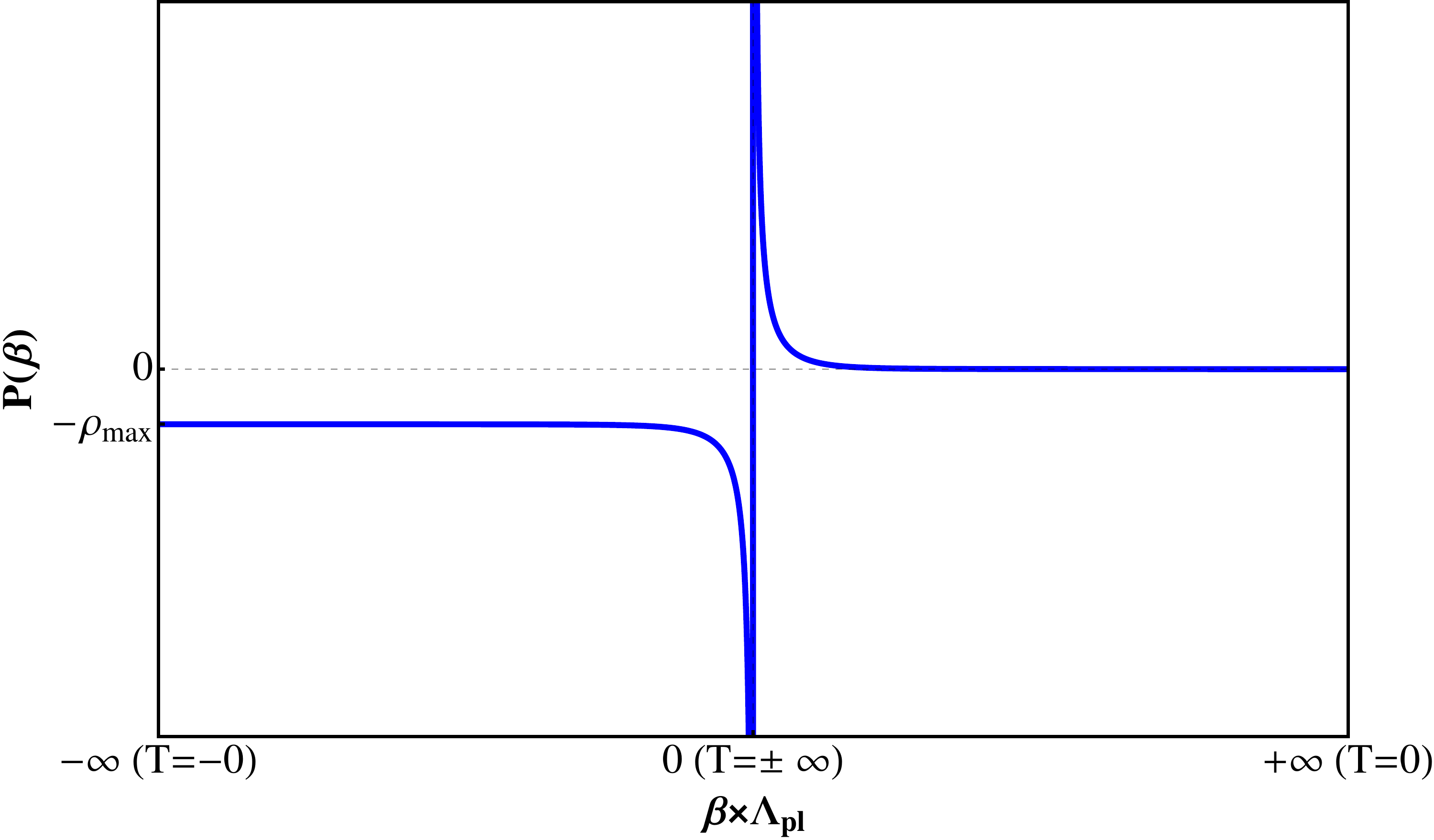,width=0.33\linewidth,clip=} 
\end{tabular}
\vspace{-0.5cm}
\caption{The dependence of the number density, the energy density, and the pressure on $\beta$.  a) The number density reaches to its maximum values at the highest temperature, $T = -0$ ({\it i.e.} $\beta = -\infty$). Then, it decreases as $T$ does. b) The energy density is maximum at the highest temperature. The energy density also decreases as $T$ does. c) The values of the pressure asymptotically approach to the constant values at both the highest and the lowest temperature ($P(T=-0) = -\rho$, $P(T=0) = 0$). Thus, NATON acts like the cosmological constant at $T = -0$ and behaves like the matter at $T = 0$.  } \label{fig-2}
\vspace{1cm}
\end{figure}
%%%%%%%%%%%%%%%%%%%%%%%%%%%%%%%%%%%%%%%
The behaviors of $n$, $\rho$, and $P$ as a function of $\beta$ are depicted in Fig.\ref{fig-2}. In the left panel, the number density of NATON dependence on the $\beta$ is described. As the temperature reaches the maximum , $T = -0$ ({\it i.e.} $\beta = -\infty$), $n$ reaches to its maximum value $n_{\max}$. When the NATON is in thermal equilibrium with the positive temperature component at $\beta = 0$, the system obtains the maximum entropy and its number density becomes half of its maximum value. If the temperature of NATON drops and reaches to zero, then there exists no more NATON. In the middle panel of Fig.\ref{fig-2}, we show the behavior of $\rho$ as a function of $\beta$.  Its behavior is almost the same as that of $n$. The energy density becomes the maximum at the highest temperature and it goes to zero at the zero temperature. The pressure of a NATON is described in the right panel of the same figure. Interestingly its effective pressure becomes negative with the absolute magnitude as $\rho_{\text{max}}$ when the temperature becomes negative zero. Thus, it causes the accelerating expansion of the Universe when NATON is dominated. After the thermal contact with the positive temperature component, NATON becomes cooler and its pressure becomes zero. Thus, NATON behaves like a matter after the thermal equilibrium with PAT. One should be noticed that the above results are only true for the relativistic limits. When the temperature drops as low as the mass of NATON, then one will not be able to use this result.   

\subsection{Mini inflation and graceful exit due to NATON}
\label{ssec:miniinfNATON}

The Big Bang provides the upper limit of energy where our classical still holds. Thus, at the beginning of the Universe, the Universe might have a chance to have a period dominated by NATON. As we show in the previous subsection \ref{ssec:rhoPNATON}, the mass of NATON should be much smaller than the Planck mass in order to have enough density. In the current laboratory observation, the duration of the existence of NAT system is about microsecond and that is why it is difficult to be observed in a daily life experience. However, the typical time scale around Big Bang is $10^{-43}$ second and it is sufficient time for NATON to contribute to the dynamics of the early Universe right after Big Bang. Especially, NATON might cause mini inflation prior to the standard one. This short inflationary epoch will solve the initial condition problems of standard inflation models. Also, this mini inflationary epoch can be naturally terminated due to the thermal contact of NATON with the radiation (or matter). It is well known that the thermal contact between the NAT system with the system of the PAT always produces a system with final PAT \cite{Puglisi_ea17}. Let us briefly review this. Initially, there are both NAT and PAT systems and the total entropy of this system is given by 
\begin{align}
S_{I} &= S^{\scaleto{\NAT\mathstrut}{3.5pt}} \left(E_{\scaleto{\NAT\mathstrut}{3.5pt}}^\text{I} \right) + S^{\scaleto{\PAT\mathstrut}{3.5pt}} \left(E_{\scaleto{\PAT\mathstrut}{3.5pt}}^\text{I} \right) \, , \label{SIP}
\end{align}
while, after the thermal contact, it becomes 
\begin{align}
S_{\text{F}} &= S^{\scaleto{\NAT\mathstrut}{3.5pt}} \left(E_{\scaleto{\NAT\mathstrut}{3.5pt}}^\text{F} \right) + S^{\scaleto{\PAT\mathstrut}{3.5pt}} \left(E_{\scaleto{\PAT\mathstrut}{3.5pt}}^\text{F} \right) \, , \label{SFP}
\end{align}
where $E_{\scaleto{\NAT\mathstrut}{3.5pt}}^\text{I} + E_{\scaleto{\PAT\mathstrut}{3.5pt}}^\text{I} = E_{\scaleto{\NAT\mathstrut}{3.5pt}}^\text{F} + E_{\scaleto{\PAT\mathstrut}{3.5pt}}^\text{F}$. The final energy of NAT system, $E_{\scaleto{\NAT\mathstrut}{3.5pt}}^\text{F}$ is determined by  the equilibrium condition that $S_{\text{F}}$ takes the maximum possible value 
\begin{align}
\beta_{\scaleto{\NAT\mathstrut}{3.5pt}}^\text{F}  &\equiv \frac{\partial  S^{\scaleto{\NAT\mathstrut}{3.5pt}} \left(E_{\scaleto{\NAT\mathstrut}{3.5pt}}^\text{F} \right)}{\partial E_{\scaleto{\NAT\mathstrut}{3.5pt}}^\text{F}} = \frac{\partial  S^{\scaleto{\PAT\mathstrut}{3.5pt}}\left(E_{\scaleto{\NAT\mathstrut}{3.5pt}}^\text{F} \right)}{\partial E_{\scaleto{\PAT\mathstrut}{3.5pt}}^\text{F}} \equiv \beta_{\scaleto{\PAT\mathstrut}{3.5pt}}^\text{F}  \label{betaNPAT}
\end{align}
Since $\beta_{\scaleto{\PAT\mathstrut}{3.5pt}}$ is positive for every value of $E_{\scaleto{\PAT\mathstrut}{3.5pt}}$, the final common temperature must also be positive.This shows that the temperature of the NATON becomes positive after it thermally contacted with a PAT system. Thus, NATON scenario will not suffer from the graceful exit problem. The NATON causes the early mini inflation prior to standard inflation and becomes the matter component after it is in thermal equilibrium with other components. In section \ref{sec:ICsol}, we suggest the existence of NATON as a possible solution for the initial condition problems of the standard inflation model which is briefly introduce in \ref{sec:ICpro}.     

\section{A brief review of initial condition problems of inflation models}
\label{sec:ICpro}
Even though the inflation models are generally accepted as a successful mechanism to supplement the drawback of the Big Bang model, the naturalness of its initial condition is still under debate \cite{Goldwirth_92}. These problems have been investigated for the model with curvature term, the anisotropic universes, and an inhomogeneous universe. The effects of the kinetic term and of the background curvature and the influence of the anisotropy and of perturbations around a cosmological constant on the onset of inflation are relatively easy to consider and are well understood. However, the initial inhomogeneity of the inflaton is the most serious problem and still in debate \cite{Mazenko_ea85,Linde_85, Goldwirth_92, Knox_ea93, Iguchi_ea97, Bolejko_ea10, Alho_ea10, Ijjas_ea13, Ijjas_ea14, Guth_ea14, Linde_14, Easther_ea14, Dalianis_ea15, Carrasco_ea15, Kaya_16, Clough_ea16, Brandenberger_16, Bastero-Gil_ea16, Bagchi_ea18, Linde_18, Mishra_ea18}. This problem can be separated as either an energy balance or a dynamical effect. The former is related to the magnitude of the gradient term of the inflaton compared to that of potential term. The latter indicates the back-reaction effects of the initial inhomogeneities on the background which might destroy the homogeneous background itself preventing the onset of inflation. We briefly summarize these problems in the following subsections \ref{ssec:ICproEB} and \ref{ssec:ICproDE}.

\subsection{Initial condition problem as a energy balance}
\label{ssec:ICproEB}

The effective energy density and the pressure of the inflaton field can be written as
\begin{align}
\rho_{\phi} &= \frac{1}{2} \left( \hbar \frac{\partial \phi}{\partial t} \right)^2 + V(\phi) + \frac{1}{2a^2} \left( \hbar c \nabla \phi \right)^2 \equiv \rho_{\phi}^{\text{hom}} + \rho_{\phi}^{\text{inh}}\, , \label{rhophi} \\
P_{\phi} &= \frac{1}{2} \left( \hbar \frac{\partial \phi}{\partial t} \right)^2 - V(\phi) - \frac{1}{6a^2} \left( \hbar c \nabla \phi\right)^2 \equiv P_{\phi}^{\text{hom}} + P_{\phi}^{\text{inh}}\, , \label{Pphi} \\
\end{align}
where $(\nabla \phi)^2 = \frac{\partial \phi}{\partial x_{i}} \frac{\partial \phi}{\partial x^{i}}$ and $x_i$ is the comoving distance which is related to the physical distance $r_i$ and the scale factor $a$, as $r_i (t) = a(t) x_i(t)$. In the standard inflation models, one ignores the gradient term $\frac{1}{2a^2} (\hbar c \nabla \phi)^2 \equiv \rho_{\phi}^{\text{inh}}$ by assuming the homogeneous background of the Universe at the beginning of the inflation. For example, the Friedmann equation for the chaotic inflation model is given by
\begin{align}
H_{I}^2 &\simeq \frac{8 \pi G}{3 c^2} V(\phi) = \frac{4 \pi}{3 c^2} \frac{\hbar c}{m_{\scaleto{\Pl\mathstrut}{5pt}}^2} \left( m_{\phi} c^2 \phi \right)^2 = \frac{4\pi}{3} \frac{m_{\phi}^2}{m_{\scaleto{\Pl\mathstrut}{5pt}}^2} \left( \hbar c \phi^2 \right) c^2 \equiv \frac{m_{\phi}^2}{m_{\scaleto{\Pl\mathstrut}{5pt}}^2} \frac{l_{\scaleto{\Pl\mathstrut}{5pt}}^3}{L^3} \left( \frac{c}{l_{\Pl}} \right)^2 \equiv \frac{1}{2} \frac{m_{\phi}^2}{m_{\scaleto{\Pl\mathstrut}{5pt}}^2} \frac{l_{\scaleto{\Pl\mathstrut}{5pt}}^3}{L^3} H_{\scaleto{\Pl\mathstrut}{5pt}}^2 \label{HI} \, , \\
\text{where} & \,\,\, G \equiv \frac{\hbar c}{m_{\scaleto{\Pl\mathstrut}{5pt}}^2} \, , \, H_{\scaleto{\Pl\mathstrut}{5pt}} \equiv \frac{\sqrt{2}  c}{l_{\scaleto{\Pl\mathstrut}{5pt}}} \equiv \frac{\sqrt{2}}{t_{\scaleto{\Pl\mathstrut}{5pt}}} \, , \, L^{-3} \equiv \frac{4\pi}{3} m_{\scaleto{\Pl\mathstrut}{5pt}} c^2 \phi^2  \, . \nonumber 
\end{align}
The above equation \eqref{HI} shows the ratio of the Hubble radius between Planck and inflation
\begin{align}
\frac{H_{I}}{H_{\scaleto{\Pl\mathstrut}{5pt}}} &= \sqrt{\frac{1}{2}} \frac{m_{\phi}}{m_{\scaleto{\Pl\mathstrut}{5pt}}} \left( \frac{l_{\scaleto{\Pl\mathstrut}{5pt}}}{L} \right)^{3/2} \label{HIoHPl} \, .
\end{align} 
In chaotic inflation models, $\phi = (\Lambda_{\Pl} l_{\Pl}^3)^{-1/2}$ ({\it i.e.} in natural units $\phi = m_{\Pl}$).  However, the effects of inhomogeneity might be critical and the large inhomogeneity can prevent inflation. To obtain the accelerating expansion, the potential should be a dominant component compared to both the kinetic and the gradient terms. In the usual slow-roll condition, the former condition is satisfied. Thus, one needs to make sure the condition for  
\begin{align}
V(\phi) > \rho_{\phi}^{\text{inh}} \,\,\, \longrightarrow \,\,\, \frac{3c^2}{8\pi G} H_{I}^2 > \frac{\left( \hbar c \right)^2}{2a_{I}^2}  \left( \nabla \phi_{I} \right)^2  \,\,\, \longrightarrow \,\,\, a_{I} \Delta_{I} > \sqrt{\frac{4\pi}{3}} l_{\Pl}^{3/2} \sqrt{ m_{\scaleto{\Pl\mathstrut}{5pt}} c^2 \phi^2} c \equiv \left( \frac{l_{\Pl}}{L_{I}} \right)^{3/2} \frac{c}{H_I} \label{ICDelta} \, ,
 \end{align} 
where $\Delta_{I}$ is the comoving wavelength of the initial perturbation of inflaton. This means that the physical length at the beginning of the inflation $a_{I} \Delta_{I}$ should be larger than a few Hubble radius at the same epoch $c/H_{I}$ because ${\cal O} (L_{I}) \simeq {\cal O} (l_{\Pl})$ in the chaotic inflation models. This means that the pre-inflationary universe should be quite homogeneous. Otherwise, a large initial inhomogeneity might prevent inflation itself. 

\subsection{Initial condition problem as a dynamical effect}
\label{ssec:ICproDE}

The dynamical effect of the gradient term might influence the evolution of inflaton and which might prevent inflation. This dynamical effect can be understood in the following way. There exists a region where the scalar field is homogeneous and suitable for inflation. There is also a sharp change in the inflaton outside of this region and it is improper for the inflation \cite{Mazenko_ea85}. The interior region inflates with its exponentially growing volume. The perturbation of the scalar field moves inwards and the surface of this region contracts. In order that the inwardly moving surface does not catch up with the expanding interior, it is required that its initial size must be sufficiently large. If the surface contracts at the speed of light, the inflation will be continued if the interior is larger than the initial horizon. It means that the initial homogeneity on a scale at least comparable to the initial horizon, $a_{I} \Delta_{I} > c H_{I}^{-1}$, is required independently of the specific inflationary model. This argument can also be used to obtain an upper limit on the required size of the homogeneous region. If the scalar field is suitable for inflation in a region with an initial size of $3 c H_{I}^{-1}$. The perturbation from the exterior will propagate inwards when the interior inflates. During one Hubble time, the perturbation propagates at most a distance of $c H_{I}^{-1}$. At the same time, the innermost region will grow by a factor of $e \simeq 3$. After one Hubble time, the size of an inner region becomes $ 3 c H_{I}^{-1}$ which is suitable for inflation. Thus, the innermost region will keep inflating in spite of the perturbation on the boundary. In this case, the inflating region will not grow and its size will always remain as $3 c H_{I}^{-1}$. If the initial size is slightly larger, then the inner region will grow and expand. This shows that $a_{I} \Delta_{I} > 3 c H_{I}^{-1}$ is a sufficient condition for the onset of inflation.

\section{A solution for initial condition problems of inflation models}
\label{sec:ICsol}

As we show in the previous section \ref{sec:ICpro}, the main problem of the initial condition on the onset of the inflation stems from the fact that the physical size when inflation starts should be bigger than that of Hubble radius during the inflation. This problem is similar to a horizon problem of standard Big Bang which can be cured by introducing an inflationary expanding epoch. In the standard inflation models, one assumes that the universe is dominated by a radiation (or matter) component during the period between Big Bang and the inflation epoch. In this case, the Hubble radius continuously increases during this epoch and the Hubble radius at the inflation is larger than that of before the inflation. This causes the IC problems in the standard inflation models.    

\subsection{Mini inflation due to NATON}
\label{ssec:mininf}

\begin{figure}[h]
\centering
\vspace{1cm}
\epsfig{file=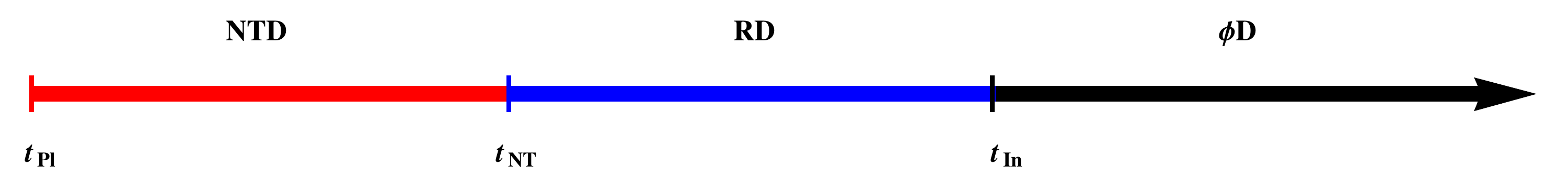,width=0.8\cw}
\caption{This shows the modified timeline of the early Universe including a NATON dominated epoch. Right after the Big Bang, the Universe is supposed to be dominated by a NATON. During this period from $t_{\Pl}$ to $t_{\NT}$, the Universe experience mini inflation. After a NATON becomes in thermally equilibrium with radiation at $t_{\NT}$, the dominant component of the Universe is either radiation or matter. Thus, the expansion of the Universe was decelerating prior to standard inflation. And the standard inflation is ignited at $t_{\text{IN}}$. NTD, RD, $\phi$D denote NATON dominated, radiation dominated, and inflaton dominated epochs, respectively. } \label{fig-3}
\vspace{1cm}
\end{figure}

In this subsection, we explicitly show the modified expansion history of the Universe due to the existence of the NATON.  Traditionally, one assumes that the Universe is dominated by radiation before the standard inflation starts. In section \ref{sec:NAT}, however, we show that the existence of the upper limit on the energy scale can naturally produce the system of NAT. Thus, we introduce the possibility of the existence of a NATON between the Big Bang and radiation dominated epoch prior to the standard inflationary epoch. As we show in subsection \ref{ssec:rhoPNATON}, the NATON behaves like the cosmological constant $P = -\rho$ right after the Big Bang before it thermally contacts with the radiation component. Thus, it provides the de Sitter expansion of the early Universe prior to the usual standard inflation
\begin{align}
a_{\scaleto{\NT\mathstrut}{5pt}}(t) &= a_{\scaleto{\Pl\mathstrut}{5pt}} \exp \left[ \, \sqrt{\frac{ \rho_{\scaleto{\NT\mathstrut}{5pt}}}{\rho_{\scaleto{\Pl\mathstrut}{5pt}}}}  \, H{\scaleto{\Pl\mathstrut}{5pt}} \left( t - t_{\scaleto{\Pl\mathstrut}{5pt}} \right) \right] 
= a_{\scaleto{\Pl\mathstrut}{5pt}} \exp \left[ \, \sqrt{\frac{ 2 \rho_{\scaleto{\NT\mathstrut}{5pt}}}{\rho_{\scaleto{\Pl\mathstrut}{5pt}}}}  \, \left( \frac{t - t_{\scaleto{\Pl\mathstrut}{5pt}}}{t_{\scaleto{\Pl\mathstrut}{5pt}}} \right) \right] \nonumber \\ 
&= a_{\scaleto{\Pl\mathstrut}{5pt}} \exp \left[ \, \sqrt{\frac{g }{6\pi}}  \, \left( \frac{t - t_{\scaleto{\Pl\mathstrut}{5pt}}}{t_{\scaleto{\Pl\mathstrut}{5pt}}} \right) \right]  \, , \label{aNTt}
\end{align}
where we use equation \eqref{rhoTmumax} and $\rho_{\Pl} = 3 m_{\Pl} /(4 \pi l_{\Pl}^3)$. We also use the fact that $\rho = \rho_{\max}$ at $T \simeq -0$ ({\it i.e.} $\beta \simeq -\infty$). In Fig.\ref{fig-3}, we describe the modification of the timeline of the Universe when we include the NATON in our traditional Big Bang model. The Planck time $t_{\Pl}$ indicates the beginning of our classical universe and it corresponds to the beginning of the NATON dominated epoch. The Universe is dominated by a NATON between $t_{\Pl}$ and $t_{\NT}$. Thus, the universe expanded at an accelerating rate given in equation \eqref{aNTt} during this period. When a NATON couples with radiation at $t_{\NT}$, it becomes like a matter component and terminates the mini inflation. During this period the Hubble radius is decreasing and this is the essential point of this model to solve the initial condition problems of the standard inflation models. The detail discussion for this is followed in the next subsection \ref{ssec:solIC}. Radiation (or matter) component dominates the Universe between $t_{\NT}$ and $t_{\text{In}}$. During this period, the Hubble radius is increasing and it is the main source of the IC problems of standard inflation models. $t_{\text{In}}$ denotes the epoch of the onset of the standard inflationary epoch. 

%%%%%%%%%%%%%%%%%%%%%%%%%%%%%%%%%%%%%%%
%%%%%%%%%%%%%%%%%%%%%%%%%%%%%%%%%%%%%%%

\subsection{Solution for initial condition problems}
\label{ssec:solIC}
We quickly review the Hubble radius before explaining how NATON can solve IC problems. The physical distance $r(t)$ and the comoving distance $\Delta$ are related to each other by the scale factor $a(t)$ as $r(t) = a(t) \Delta$. By differentiating the above relation and replacing physical velocity with the speed of light $c$, one obtains the definition of the Hubble radius $R_{\text{H}}$ 
\begin{align}
R_{\text{H}} &\equiv \frac{c}{a H} \, . \label{RH} 
\end{align}
Thus, the Hubble radius is the comoving distance over which particles can travel in the course of one expansion time. If the comoving separation of two particles is larger than the Hubble radius at some epoch, then the particles are causally disconnected with each other at that epoch. In section \ref{sec:ICpro}, we show that sufficient condition is $\Delta_{I} > 3c/(aH)$ In order to solve the IC problems. The mini inflation due to the NATON can satisfy this condition easily. We investigate this in both qualitatively and quantitatively.  

%%%%%%%%%%%%%%%%%%%%%%%%%%%%%%%%%%%%%%%
\begin{figure}[h]
\centering
\vspace{1cm}
\epsfig{file=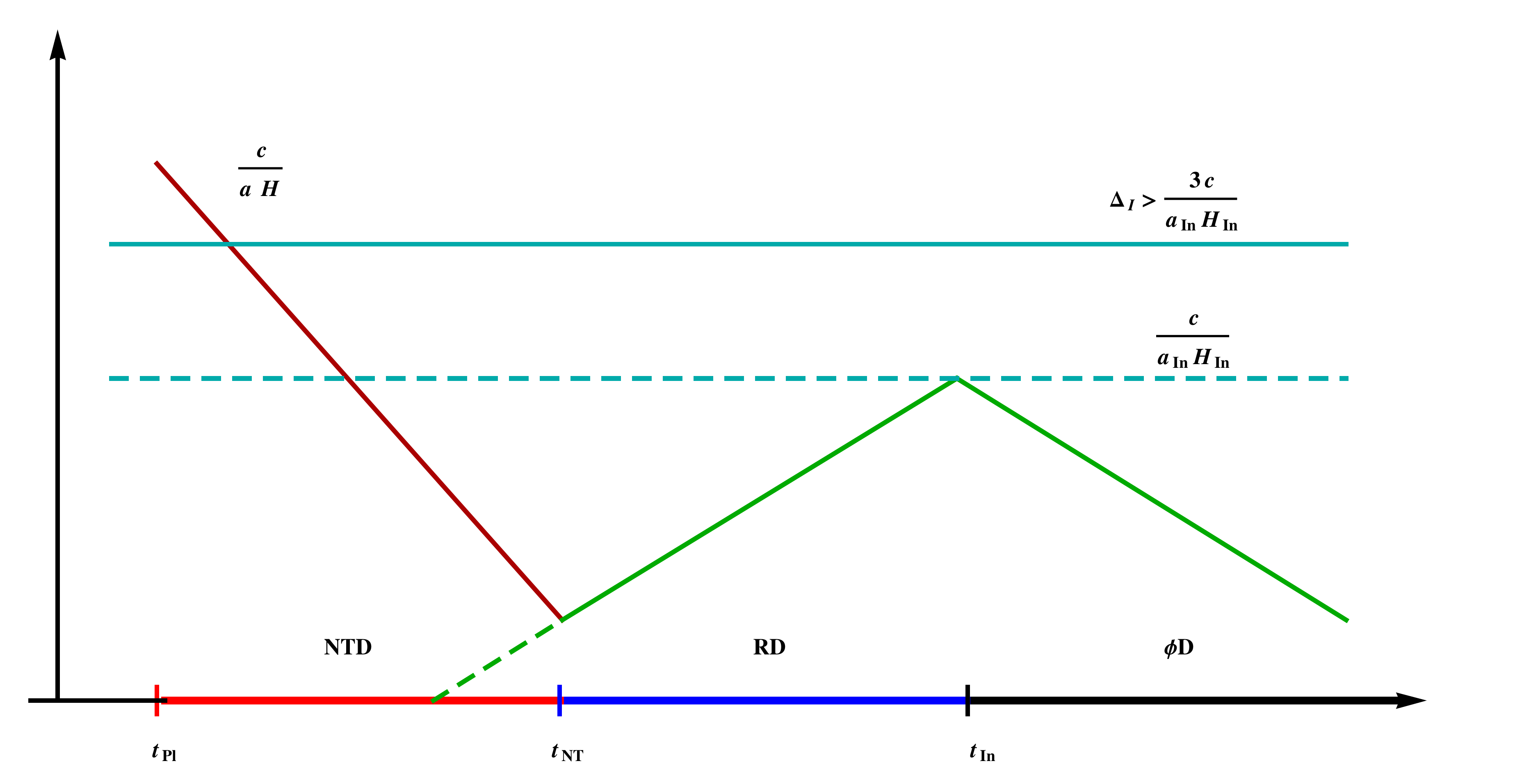,width=0.8\cw}
\caption{The evolution of the Hubble radius $c/(aH)$ as a function of time. In NATON model, $R_{\text{H}}$ decreases during a NATON dominated epoch because the expansion of the Universe is accelerating during this epoch. After $t_{\NT}$, the Universe is under a decelerating expansion during the radiation dominated epoch and $R_{\text{H}}$ increases and reaches its values $c/(a_{\text{In}} H_{\text{In}})$ at $t_{\text{In}}$. The expansion of the Universe is accelerating again at $t_{\text{In}}$ to give decreasing $R_{\text{H}}$ again.} \label{fig-4}
\vspace{1cm}
\end{figure}
%%%%%%%%%%%%%%%%%%%%%%%%%%%%%%%%%%%%%%%
It is easy to understand qualitatively how NATON which causes mini inflation can solve the IC problems from figure \ref{fig-4}. In this figure, we show the evolution of the Hubble radius at the early Universe. The solid inclined lines correspond $R_{\text{H}}$ at different epoch. It decreases from the Big Bang to right before the onset of radiation dominated epoch. Thus, the earlier the $t_{\Pl}$ relative to $t_{\NT}$, the larger the $R_{\text{H}}(t_{\Pl})$. After $t_{\NT}$, $R_{\text{H}}$ increases until it reaches to its intermediated maximum value $R_{\text{H}}(t_{\text{In}})$. If there exists no NATON dominated epoch, then the Hubble radius keeps increasing as time increases in the radiation dominated epoch. This is indicated as the dotted inclined line. When the inflation is ignited,  $R_{\text{H}}$ decreases again and this solves the standard horizon problems, flatness problems, and etc. As shown in this figure, the existence of mini inflationary era prior to the standard inflationary epoch makes it possible to have the comoving distance which is bigger than $3 c /(a_{\text{In}} H_{\text{In}})$ at the onset of the standard inflation. This solves the IC problems and makes standard inflation models free from the fine-tuning problem.  

Now we are ready to investigate the solution of IC problems quantitatively. The scale factors in both NATON and radiation dominated epochs are given by
\begin{align}
a(t) &= a_{\scaleto{\NT\mathstrut}{5pt}} \exp \left[ \, H_{\scaleto{\NT\mathstrut}{5pt}} \left( t - t_{\scaleto{\NT\mathstrut}{5pt}} \right) \right] \,\,\, \text{when} \,\,\, t_{\scaleto{\Pl\mathstrut}{5pt}} \leq t \leq t_{\scaleto{\NT\mathstrut}{5pt}} \, , \label{aNT} \\
a(t) &= a_{\scaleto{\NT\mathstrut}{5pt}} \sqrt{ 2 H_{\scaleto{\NT\mathstrut}{5pt}} \left( t - t_{\scaleto{\NT\mathstrut}{5pt}} \right) + 1} \,\,\, \text{when} \,\,\, t_{\scaleto{\NT\mathstrut}{5pt}} \leq t \leq t_{\scaleto{\text{In}\mathstrut}{5pt}} \, , \label{aIn} \\
\end{align}
where $H_{\scaleto{\NT\mathstrut}{5pt}} = \sqrt{g/(6\pi)} H_{\scaleto{\Pl\mathstrut}{5pt}} = \sqrt{g/(3\pi)} t_{\scaleto{\Pl\mathstrut}{5pt}} ^{-1}$. From the above equations \eqref{aNT}-\eqref{aIn}, one can also obtain the inverse Hubble radius divided by the speed of light $c$ 
\begin{align}
a(t) H(t) &=  a_{\scaleto{\NT\mathstrut}{5pt}}  H_{\scaleto{\NT\mathstrut}{5pt}} \exp \left[ \, H_{\scaleto{\NT\mathstrut}{5pt}} \left( t - t_{\scaleto{\NT\mathstrut}{5pt}} \right) \right] \,\,\, \text{when} \,\,\, t_{\scaleto{\Pl\mathstrut}{5pt}} \leq t \leq t_{\scaleto{\NT\mathstrut}{5pt}} \, , \label{aHt} \\
a(t_{\scaleto{\text{In}\mathstrut}{5pt}})  H(t_{\scaleto{\text{In}\mathstrut}{5pt}} ) &=  a_{\scaleto{\NT\mathstrut}{5pt}}  H_{\scaleto{\NT\mathstrut}{5pt}} \left( 2 H_{\scaleto{\NT\mathstrut}{5pt}} \left( t_{\scaleto{\text{In}\mathstrut}{5pt}} - t_{\scaleto{\NT\mathstrut}{5pt}} \right) + 1 \right)^{-1/2} \,\,\, \text{when} \,\,\, t = t_{\scaleto{\NT\mathstrut}{5pt}} \, . \label{aHIn} \\
\end{align}
In order to solve the IC problem, we need a condition that NATON dominated epoch should be last at least until at $t_{\ast}$ 
\begin{align}
\frac{c}{a(t_{\ast}) H (t_{\ast}) } \geq 3 \frac{c}{a(t_{\scaleto{\text{In}\mathstrut}{5pt}})  H(t_{\scaleto{\text{In}\mathstrut}{5pt}} ) } \label{Solcon} \, .
\end{align}
One can insert Eqs.\eqref{aHt} and \eqref{aHIn} into equation \eqref{Solcon} to obtain
\begin{align}
& \exp \left[ \, -2 \sqrt{\frac{g}{3\pi}} \left( \frac{t_{\ast}}{t_{\scaleto{\Pl\mathstrut}{5pt}} } - \frac{t_{\scaleto{\NT\mathstrut}{5pt}}}{t_{\scaleto{\Pl\mathstrut}{5pt}} }  \right) \right] \geq 9 \left( 2\sqrt{\frac{g}{3\pi}} \left( \frac{t_{\scaleto{\text{In}\mathstrut}{5pt}}}{t_{\scaleto{\Pl\mathstrut}{5pt}} }  - \frac{t_{\scaleto{\NT\mathstrut}{5pt}}}{t_{\scaleto{\Pl\mathstrut}{5pt}} }  \right) + 1 \right) \, , \label{tast} \\
& \rightarrow  \,\,\, t_{\scaleto{\NT\mathstrut}{5pt}} \geq \sqrt{\frac{3 \pi}{4 g}} \ln \left[ \sqrt{\frac{108 g}{\pi}} \frac{t_{\scaleto{\text{In}\mathstrut}{5pt}}}{t_{\scaleto{\Pl\mathstrut}{5pt}} } \right] t_{\scaleto{\Pl\mathstrut}{5pt}} \simeq 34 \, t_{\scaleto{\Pl\mathstrut}{5pt}} \nonumber \, ,
\end{align}
where we use the fact that $\frac{t_{\scaleto{\text{In}\mathstrut}{5pt}}}{t_{\scaleto{\Pl\mathstrut}{5pt}} }  >> \frac{t_{\scaleto{\NT\mathstrut}{5pt}}}{t_{\scaleto{\Pl\mathstrut}{5pt}} }   >>  \frac{t_{\ast}}{t_{\scaleto{\Pl\mathstrut}{5pt}} }$. Thus, the standard inflation models have the enough homogeneous region as long as NATON dominates the Universe until roughly 30 times longer than the Planck time. And this is quite plausible.  

\section{DISCUSSIONS AND CONCLUSIONS}
\label{sec:DisCon}

We show that there might be a negative absolute temperature system in the early Universe due to the natural existence of an upper bound to the energy of its system, dubbed ``NATON''. Even though its duration is very short compared to daily life experience, it is a long enough time in the early Universe to give its effect to the dynamics of the Universe. It provides negative pressure and it causes the accelerating expansion of the early Universe prior to the standard inflationary epoch. This mini inflation solves the initial condition problems of the standard inflation and makes it a natural theory. In order to solve the initial condition problem, NATON requires to exists until 30 times bigger than the Planck time. NATON dominated epoch is terminated by thermal contact with the positive absolute temperature system. Thus, it will not suffer the graceful exit problem either. 

There have been several nice investigations related the NAT with dark energy \cite{Braun_ea13,Vieira_ea16,Saha_ea18}. However, we think this might be more difficult to be realized than the NATON model because of two reasons. The first is related to the existence of the upper bound in the energy.  In order to realize the NAT, there should be the upper bound but there might not be any upper energy bound in the late Universe. The second is the problem of the duration of NAT. Because if any NAT system can give effect on the dynamics of the Universe, it should last until the present age of the Universe. However, it can exist only microsecond in most case and it might not give any effect on the Universe even though it exists. There is also condensed matter experiment which claims that system mimics the expansion of the Universe \cite{Eckel_ea18}. It is interesting to see attempts to implement the evolution of the universe in the laboratory and it might be able to give us any clue about the universe.

One also might think about the observational effects of the NATON dominated epoch on cosmic microwave background or large scale structure. Previously, there have been interesting works related to the effect of radiation dominated epoch prior to the standard inflation era on cosmological observables \cite{Wang:2007ws,Destri:2009hn,Das:2014ffa,Bahrami:2015bva,Das:2015ywa,Musmarra:2017rfu,Sasaki:2018ang}. It will be interesting if one can find any effect of NATON on those observables. However, this is beyond the scope of this manuscript and this subject will be discussed in the following manuscript.  

\begin{acknowledgments}
SL is supported by Basic Science Research Program through the National Research Foundation of Korea (NRF) funded by the Ministry of Science, ICT and Future Planning (Grant No. NRF-2017R1A2B4011168).
\end{acknowledgments}

%\bibliography{lpt_refs}

\end{document}